\documentstyle[version2,preprint,aps]{revtex} 
\begin{document}                                                        
\draft                                                                  
                                                             
\begin{title}
Validity of the rigid band picture for the $t$$-$$J$ model
\end{title}

\author{R. Eder, Y. Ohta and T. Shimozato}

\begin{instit}
Department of Applied Physics, Nagoya University, Nagoya 464-01, Japan
\end{instit}

\begin{abstract}
We present an exact diagonalization study
of the doping dependence of the
single particle Green's function in 
$16$, $18$ and $20$-site clusters
of $t$$-$$J$ model. We find evidence for the
validity of the rigid-band picture starting from the
half-filled case: upon doping, the topmost states of the
quasiparticle band observed 
in the photoemisson spectrum at half-filling
cross the chemical potential and reappear as the lowermost states
of the inverse photoemission spectrum.
Features in the inverse photoemission spectra, which 
are inconsistent with the rigid band picture,
are shown to originate from the nontrivial point group
symmetry of the ground 
state with two holes, which enforces different
selection rules than at half-filling.
Deviations from rigid band behaviour which lead to the
formation of a `large Fermi surface' in the momentum distribution
are found to
occur at energies far from the chemical potential.
A Luttinger Fermi surface and
a nearest neighbor hopping band do not exist.
\end{abstract} 


A well known problem in the description of high-temperature
superconductors is the volume of the Fermi surface.
Since these systems are close to a 
metal-to-insulator transition, there arises the
following question: should one model them by a system of
quasiparticles which correspond to the doped holes and 
populate the dispersion relation calculated
for a single hole
(rigid band approximation) or should one assume that
the ground state can still be obtained by 
adiabatic continuation from the noninteracting one,
so that the Fermi surface corresponds to
a slightly less than half-filled band of noninteracting
electrons?
Based on numerical studies of the momentum distribution and
single particle spectral function for the frequently used $t$$-$$J$ 
model it has been argued\cite{StephanHorsch} that 
the single hole represents a `problem of only marginal relevance'
for the doped case:
already for two holes in clusters with $16$$-$$20$ lattice sites
(corresponding to a nominal hole concentration of $\sim$$10$ \%)
a kind of phase transition has ocurred so that both,
the Fermi surface and the quasiparticle band structure in its
neighborhood, resemble that for noninteracting particles.
In this manuscript we present evidence against this
widely accepted picture:
as far as the photoemission spectrum is concerned
the rigid band approximation (RBA)
is in fact an excellent one;
to fully understand the inverse photoemission 
spectrum one has to take into account the nontrivial 
symmetry of the two-hole ground
state, which enforces transitions into a second, symmetry-different
band of many-body states. \\
The $t$$-$$J$ model reads
\[
H =                                                    
 -t \sum_{< i,j >, \sigma}                                         
( \hat{c}_{i, \sigma}^\dagger \hat{c}_{j, \sigma}  +  H.c. )
 + J \sum_{< i,j >}(\bbox{S}_i \cdot
 \bbox{S}_j
 - \frac{n_i n_j}{4}).
\]
The $\bbox{S_i}$ are the electronic                                        
spin operators and                                                             
the sum over $<i,j>$ stands for a summation                                    
over all pairs of nearest neighbors.
The operators $\hat{c}_{i,\sigma}$
are expressed in terms of ordinary fermion                                  
operators as $c_{i,\sigma}(1-n_{i,-\sigma})$.
We study the single particle spectral function
$A_{n}(\bbox{k},\omega) = A_{n,-}(\bbox{k},-\omega) 
+ A_{n,+}(\bbox{k},\omega)$, 
where the photoemission (PES)
spectrum $A_{n,-}(\bbox{k},\omega)$ and inverse photoemission (IPES)
spectrum $A_{n,+}(\bbox{k},\omega)$ are defined as
\begin{eqnarray}
A_{n,-}(\bbox{k}, \omega) &=&
\sum_{\nu} | \langle \Psi_{\nu,n+1} | \hat{c}_{\bbox{k},\sigma}|
\Psi_{0,n} \rangle |^2
\nonumber \\
&\;&\;\;\;\delta( \omega - ( E_{\nu,n+1} - E_{ref} )),
\nonumber \\
A_{n,+} (\bbox{k}, \omega) &=&
\sum_{\nu} | \langle \Psi_{\nu,n-1} 
| \hat{c}_{\bbox{k},\sigma}^\dagger |
\Psi_{0,n} \rangle |^2
\nonumber \\
&\;&\;\;\;\delta( \omega - ( E_{\nu,n-1} - E_{ref} )).
\label{spectral}
\end{eqnarray}
Here $|\Psi_{\nu,n} \rangle$  ($E_{\nu,n}$)  
is the $\nu^{th}$ eigenstate (eigenenergy) with
$n$ holes (in particular $\nu$$=$$0$ implies the ground state)
and the reference energy $E_{ref}$ is 
chosen as $E_{0,n}$ (although other values will be advantageous
later on). All spectra were evaluated exactly by the Lanczos method.\\ 
%%%%%%%%%%%%%%%%%%%%%%%%%%%%%%%%%%%%%%%%%%%%%%%%%%%%%%%%%%%%%%%%%
Let us first recall a few constraints due to sum rules.
The momentum distribution 
$n(\bbox{k})$$=$$\langle \Psi_{0,n}|  \hat{c}_{\bbox{k},\sigma}^\dagger 
\hat{c}_{\bbox{k},\sigma}|\Psi_{0,n} \rangle$
in the $n$-hole ground state is the zero$^{th}$ moment
of $A_{n,-}(\bbox{k},\omega)$; the
zero$^{th}$ moment of $A_{n,+}(\bbox{k},\omega)$, 
is given by
$1$$-$$N/2L$$-$$n(\bbox{k})$, where 
$N$ is the number of electrons and $L$ the number of sites
(for an even number of electrons and total $z$-spin $0$).
Introducing $\bbox{Q}$$=$$(\pi,\pi)$ and
$\delta n(\bbox{k})$$=$$n(\bbox{k})$$-$$n(\bbox{k}+ \bbox{Q})$, 
the expectation value of the kinetic energy
can be written as 
\begin{equation}
\langle H_t \rangle=
\sum_{\bbox{k} \in AFBZ}
\epsilon(\bbox{k}) \delta n(\bbox{k}),
\label{kinsum}
\end{equation}
where $\epsilon(\bbox{k})$ is the free-particle energy,
and the summation over $\bbox{k}$
is restricted to the interior of the antiferromagnetic
Brillouin zone ($AFBZ$). Since $\epsilon(\bbox{k})$$<$$0$ in the
range of summation, in order to have $\langle H_t\rangle$$<$$0$
the average of $\delta n(\bbox{k})$ in the $AFBZ$ must be positive i.e.
$n(\bbox{k})$ must be larger inside the $AFBZ$ 
than outside, so that there is always a 
tendency towards the formation of a `large Fermi surface'
in $n(\bbox{k})$. To exemplify this,
Tab. \ref{table1}, compares $n(\bbox{k})$ for the
two-hole ground state of the
fully isotropic $t$$-$$J$ model and of the
$t$$-$$J_z$ model (where the transverse part of the Heisenberg
exchange is discarded) with an added staggered field.
The latter model has explicitely
broken symmetry (the ground state expectation value
of the staggered magnetization is $61.3$\% of the value for the
N\'eel state) which rigorously excludes a large
Fermi surface. In spite of this,
$n(\bbox{k})$ is almost indistinguishable
for the two models and in particular always would suggest a
large Fermi surface. This shows that use of $n(\bbox{k})$
to assign a Fermi surface\cite{StephanHorsch} 
may be quite problematic.
The key problem is, that for strongly correlated systems
the `quasiparticle peak'
near the chemical potential carries only
a small fraction of the total PES weight;
if we decompose $n(\bbox{k})$ into a component
from the coherent peak, $n_{QP}(\bbox{k})$, 
and from the integration over the incoherent continua,
$n_{inc}(\bbox{k})$, we usually have
$n_{inc}(\bbox{k}) > n_{QP}(\bbox{k})$.
Then, when only $n(\bbox{k})$ is considered,
for example a change in $n_{inc}(\bbox{k})$
(which is unrelated to low-energy physics)
may mimick a Fermi surface.
As will be seen below, this is precisely what happens
when the $t$$-$$J$ model is doped with holes.
Next, the sum rules for the spectral function enforce
that in the doped case PES 
weight (IPES weight) is concentrated inside (ouside) of
the $AFBZ$, so that not only $n(\bbox{k})$ but also the
distribution of spectral weight in $\bbox{k}$-space
is reminiscent of free electrons.\\
%%%%%%%%%%%%%%%%%%%%%%%%%%%%%%%%%%%%%%%%%%%%%%%%%%%%%%%%%%%%%%%%%
Let us next discuss what can be reasonably expected if the
RBA were to hold: a variety of diagonalization 
studies\cite{BoncaII,HaPo,Itoh} have shown
that two holes in the $t$$-$$J$ model form a bound state with
a binding energy ($E_B$$\sim$$0.8 J$$-$$J$) that is a sizeable fraction
of the single-hole bandwidth ($W$$\sim$$2J$).
The two-hole ground state thus should be modelled by a 
state of the type
$|\Phi_0\rangle = \sum_{\bbox{k}} \Delta(\bbox{k})
a_{\bbox{k},\uparrow}^\dagger a_{\bbox{k},\downarrow}^\dagger
|0\rangle$,
where $a_{\bbox{k},\sigma}^\dagger$ is the creation operator
for a hole-like `quasiparticle' in the band observed 
in $A_{0,-}(\bbox{k})$.
The wave function $\Delta(\bbox{k})$ 
may differ appreciably from zero for all quasiparticle states 
within $\sim$$E_B$ above the ground state so that a 
`Fermi surface' does not exist. In IPES we
annihilate a hole (with momentum $\bbox{k}$) so
the remaining hole should be in a state belonging to
the single hole band with momentum $-$$\bbox{k}$;
the intensity should be proportional to
the quasiparticle occupation $\tilde{n}(\bbox{k})$.
Conversely, in photoemission we add a third hole and,
neglecting all interactions except for the Pauli principle,
should observe the same spectrum as for a single hole with
the intensity of the
peaks near the Fermi energy $E_f$ being reduced
by a factor $1$$-$$\tilde{n}(\bbox{k})$.
Adding up the weights of the
peaks in $A_{2,+}(\bbox{k},\omega)$ and 
$A_{2,-}(\bbox{k},\omega)$ closest to $E_f$
we should therefore ideally obtain the weight of the
`unsplit' peak in $A_{0,-}(\bbox{k},\omega)$, a
sensitive quantitative test for the RBA.
Moreover, the $d_{x^2-y^2}$ symmetry of the two-hole 
ground state\cite{HaPo}
implies that in the $16$ and $18$ site cluster 
$\Delta(\bbox{k})$ must have a node along 
the $(1,1)$ direction and hence $\tilde{n}(\bbox{k})$$=$$0$ 
for these momenta.\\ 
%%%%%%%%%%%%%%%%%%%%%%%%%%%%%%%%%%%%%%%%%%%%%%%%%%%%%%%%%%%%%%%%%
Let us now check, in how much these expectations are borne out by 
the exact spectra. Fig. \ref{spec1}
compares $A(\bbox{k},\omega)$ in the half-filled and
two-hole ground state for all allowed momenta
in the $4$$\times$$4$ and $18$-site cluster ($(0,0)$ and 
$(\pi,\pi)$ are from the $18$-site cluster),
Fig. \ref{spec21} displays the same information for the
$20$-site cluster.
No adjustment of any kind has been performed.
In agreement with the above discussion
Fig. \ref{spec1} shows that
for momenta along the $(1,1)$ direction
(left panel) there is a striking similarity between 
the PES spectra for the doped and undoped
case, $A_{2,-}(\bbox{k},\omega)$ and $A_{0,-}(\bbox{k},\omega)$,
particularly near the Fermi energy $E_f$$\sim$$+$$1.7t$.
A band of peaks with practically identical dispersion and weight
can be clearly identified in both groups of spectra
(a possible exception is $(\frac{2\pi}{3},\frac{2\pi}{3})$;
this momentum might play a special role since
the single hole ground states at this momentum and $(\pi,\pi)$ have
total spin $S=3/2$, whereas all other single hole ground states
are dubletts\cite{HaPo}). Away from $(1,1)$
(right panel of Fig. \ref{spec1})
doping leads to a shift of weight from the PES band to
IPES peaks immediately above $E_f$:
this suggests the `split peak' situation
$\tilde{n}(\bbox{k})$$\neq$$0$.
The situation is the same for the $20$-site cluster,
Fig. \ref{spec21} shows the same similarity between 
the low energy parts of the
PES spectra for the doped and undoped case.
Again the the dominant low energy PES peaks
near the Fermi energy
(right panel) remain either unaffected by doping 
or seem to (partially) cross the chemical potential to reappear
as low energy IPES peaks. Point group symmetry
poses no significant constraint on the hypothetical pair
wave function $\Delta(\bbox{k})$ in this cluster,
(all momenta except $(0,0)$ and $(\pi,\pi)$
have low symmetry) and consequently we observe
low energy IPES weight for all
momenta where low energy PES peaks were present
at half-filling. \\
It is only at energies remote from $E_f$ that major
changes in the PES spectra occur upon doping:
for momenta around $(0,0)$ and $(\pi,\pi)$
there is a pronounced addition/depletion of
incoherent spectral weight at energies $\sim$$3t$ below 
$E_f$. The increase/decrease of the integrated spectral weight
near $(0,0)$/$(\pi,\pi)$
and the corresponding formation of a `large Fermi surface'
in $n(\bbox{k})$ upon doping therefore is clearly not 
the consequence of the phase transition-like
emerging of a nearest neighbor hopping band\cite{StephanHorsch}
in the range $2J\ll 3t$ around $E_f$
(note that the spectral weight near $E_f$
at $(0,0)$/$(\pi,\pi)$ even diminishes/increases upon doping).
Rather it
is accomplished by the reshuffling of incoherent
spectral weight deep below $E_f$, and thus is certainly 
unrelated to any low-energy physics.
As a matter of fact, the actual form of $A(\bbox{k},\omega)$
for the doped $20$-site cluster rules out the
Luttinger Fermi surface postulated\cite{StephanHorsch}
for this cluster on the basis of the momentum distribution.
Despite the fact that
$(\pi,0)$ and $(\pi/5, 3\pi/5)$ are on opposite
sides of the Luttinger Fermi surface,
the low energy peak structure in $A(\bbox{k},\omega)$
is practically identical for these momenta:
both photoemisson spectra show a pronounced low 
energy peak, indicating
that these momenta are `occupied'. The criterion
$n(\bbox{k})<1/2$ ($n(\bbox{k})>1/2$), 
employed in Ref. \cite{StephanHorsch}
to decide whether a $\bbox{k}$-point is inside (outside) the
Fermi surface, obviously
has no significance for predicting the low energy
behaviour of the spectral function.\\
%%%%%%%%%%%%%%%%%%%%%%%%%%%%%%%%%%%%%%%%%%%%%%%%%%%%%%%%%%%%%%%%%
Next, we proceed to a quantitative check of the
rigid-band approximation: we
set $E_{ref}=E_{0,2}$ in  
the photoemission spectral function at half-filling,
$A_{0,-}(\bbox{k},\omega)$ and do not invert the
sign of $\omega$,
so that we can directly compare the positions of peaks 
in this function and in
the inverse photoemission spectrum for the doped ground state,
$A_{2,+}(\bbox{k},\omega)$ (both spectra involve the single hole
subspace in their final states, so that direct comparison of states
is possible).
Then, Fig. \ref{spec2} confirms that 
in the $16$ and $18$-site cluster
the final states for the lowest IPES peaks at all momenta
off the $(1,1)$ direction (right panel) indeed
belong to the single-hole band observed in PES at half-filling
(the energies of the respective lowest peaks
agree to $10^{-10}t$, essentially the limit of the Lanczos procedure).
Fig. \ref{spec22} shows that the same holds true
in the $20$-site cluster for all momenta except
$(0,0)$ and $(\pi,\pi)$.
In complete agreement with the RBA, the lowermost
peaks of the IPES spectrum for the doped case
are thus identified as the uppermost states in the PES spectrum
for the undoped case.
It is only at higher energies
($\sim J$ above the quasiparticle states)
that there are states with appreciable weight
in the IPES spectra which have
vanishing or small weight in the PES spectrum at half-filling.
The low energy physics thus should be
completely consistent with rigid-band behaviour.\\
It should be noted, that the above result is in strong
contradiction to the 
`large Fermi surface scenario'\cite{StephanHorsch}.
This would necessitate the assumption 
that the uppermost states of the
completely filled, next-nearest neighbor hopping band
observed in PES at half-filling simultaneously belong to
a half-filled, nearest neighbor hopping 
(i.e. topologically different) band 
`observed' for the two-hole ground state. 
It would moreover require the assumption
that upon doping
a full-scale transition to a topologically different
band structure can occur, while simultaneously the states
next to the chemical potential remain unaffected and 
merely cross the Fermi level.\\
%%%%%%%%%%%%%%%%%%%%%%%%%%%%%%%%%%%%%%%%%%%%%%%%%%%%%%%%%%%%%%%%%%%%%%%%%%%
Let us now turn to a discussion 
of the IPES final states in the $16$ and $18$ site cluster
along the $(1,1)$ direction
(left panel of Fig. \ref{spec2}). Obviously,
these states are not observed
in $A_{0,-}(\bbox{k},\omega)$ so that
we seem to have found `new states', which were
`generated by doping'.
The true explanation, however, is much simpler:
for each momentum along $(1,1)$ the little group
comprises the reflection by a plane along the
$(1,1)$ direction (which we denote by $T_m$). 
The ground state at half-filling is
totally symmetric, and consequently even under $T_m$,
whereas the two-hole ground state has
$d_{x^2-y^2}$ symmetry and
hence is odd under $T_m$. Consequently, for any single hole
state $|\Psi_{\nu,1} (\bbox{k})\rangle$ the matrix element
$\langle \Psi_{\nu,1} (\bbox{k}) | \hat{c}_{\bbox{k},\sigma} 
| \Psi_{0,0}\rangle$ 
($\langle \Psi_{\nu,1} (\bbox{k}) | 
\hat{c}_{\bbox{k},-\sigma}^\dagger 
| \Psi_{0,2}\rangle$) is different from zero only if
$|\Psi_{\nu,1} (\bbox{k})\rangle$ is even (odd) under $T_m$.
The appearance of the `new states'
thus is simply the consequence of a group theoretical
selection rule\cite{EderWrobel}.\\
%%%%%%%%%%%%%%%%%%%%%%%%%%%%%%%%%%%%%%%%%%%%%%%%%%%%%%%%%%%%%%%%%
We thus have to explain single-hole states with an odd parity
under $T_m$ (for the highly symmetric $\bbox{k}$-points
$(0,0)$ and $(\pi,\pi)$ they must have the full
$d_{x^2-y^2}$-symmetry).
The totally symmetric single hole states observed in PES
at half-filling can be understood in terms of
the string picture\cite{Shr1,Trugman,Maekawa,I},  
where the hole is assumed
to be dressed by a cloud of spin defects.
We thus adopt the hypothesis\cite{EderWrobel} 
that the odd-parity states in question
are similar in nature, but that the cloud of spin defects 
surrounding the hole has a nontrivial symmetry. Hence we define
\[
d_{j,\uparrow} =
S_j^- \hat{c}_{j + \hat{x},\downarrow} +
S_j^- \hat{c}_{j - \hat{x},\downarrow} -
S_j^- \hat{c}_{j + \hat{y},\downarrow} -
S_j^- \hat{c}_{j - \hat{y},\downarrow},
\]
where $j+\hat{x}$ denotes the nearest neighbor of $j$ in positive
$x$-direction etc.
When acting on the N\'eel state,
$d_{j,\uparrow}$ generates the four strings of length
$1$ beginning at site $j$.
Their relative signs makes sure that the resulting
state has $d_{x^2-y^2}$-symmetry under rotations around
$j$. A coherent superposition of such operators with momentum
along $(1,1)$ consequently creates a state with the desired
tranformation properties and we assume, that the states in
question can be described by such a wave function.
Then, for momenta along $(1,1)$, 
in $A_{0,-}(\bbox{k},\omega)$ we replace
$\hat{c}_{\bbox{k},\sigma}$
by the Fourier transform of $d_{j,\sigma}$. We denote the
resulting function by $\tilde{A}_{0,-}(\bbox{k},\omega)$ 
and it is shown in Fig. \ref{spec3}. Again we choose
$E_{ref}=E_{0,2}$ to 
faciliate comparison with the IPES spectra for the two-hole
ground state.
$\tilde{A}_{0,-}(\bbox{k},\omega)$ is similar in
character to $A_{0,-}(\bbox{k},\omega)$ in that there is
a `quasiparticle peak' at the bottom of an
incoherent continuum.  We conclude that
there indeed exists a band of many-body
states, where a hole surrounded by a spin defect cloud with
intrinsic $d_{x^2 -y^2}$-symmetry propagates coherently.
Next, the dominant peaks in the IPES spectra 
along $(1,1)$ fall precisely into this band
(the energies of the respective lowest peaks
agree to an accuracy of $10^{-10}$),
so that we have clarified their nature.\\
As a more quantitative check of the RBA
we next consider the weight of the peaks near $E_f$:
Fig. \ref{poles}
compares the weight of the peak at $(\pi/2,\pi/2)$ in
$A_{0,-}(\bbox{k},\omega)$ 
(where it equals the weight at $(\pi,0)$)
and $A_{2,-}(\bbox{k},\omega)$ 
as well as the sum of the weights of the lowest 
peak in $A_{2,+}(\bbox{k},\omega)$ 
and highest peak in $A_{2,-}(\bbox{k},-\omega)$ 
at $(\pi,0)$. The RBA predicts all three quantities
to be equal and Fig. \ref{poles} shows that they indeed
agree remarkably well over a wide range of $t/J$.\\
%%%%%%%%%%%%%%%%%%%%%%%%%%%%%%%%%%%%%%%%%%%%%%%%%%%%%%%%%%%%%%%%%
Finally, let us discuss the `band structure' near $E_f$,
summarized in Fig. \ref{band} for the $16$ and $18$-site cluster.
In $A_{2,-}(\bbox{k},\omega)$ we observe very much the same band as 
in $A_{0,-}(\bbox{k},\omega)$, 
the dispersion of the peaks close to $E_f$
being practically identical to that for the half-filled band.
%%%%%%%%%%%%%%%%%%%%%%%%%%%%%%%%%%%%%%%%%%%%%%%%%%%%%%%%%%%%%%%%%%%%%%%%%%%
For the doped case, there is moreover an obvious correlation
between the peak intensity and the distance from $E_F$
as one would expect it for a Fermi liquid:
comparison of Figs. \ref{spec1} and \ref{band}
shows that sharp peaks exist for those momenta
which are closest to the Fermi energy in their respective
clusters, weak or diffuse peaks are seen for momenta
which are more distant from $E_F$
(for momenta ouside the AFBZ there is in addition
a depletion of intensity over the whole width
of the spectra, as necessitated by the kinetic-energy 
sum rule (\ref{kinsum})). The same overall trend can also be seen
in the $20$-site cluster (Fig. \ref{spec21}).
%%%%%%%%%%%%%%%%%%%%%%%%%%%%%%%%%%%%%%%%%%%%%%%%%%%%%%%%%%%%%%%%%%%%%%%%%%%
In IPES for the $16$ and $18$ site cluster, 
the situation is more complicated due to 
the novel selection rule for momenta along the $(1,1)$ direction.
Away from this line
we observe in $A_{2,+}(\bbox{k},\omega)$ a portion of the
band seen in $A_{0,-}(\bbox{k},\omega)$
(as is the case for all momenta in the $20$-site cluster,
where no selection rule exists).
In the sense of the RBA these states 
have partially crossed the Fermi energy;
due to the interaction between the holes, however, there is no
Fermi surface but rather a zone of partially occupied
momenta (indicated by the box in Fig. \ref{band})
where the quasiparticle peaks are split 
between PES and IPES.
For IPES along $(1,1)$-direction in the $16$ and $18$-site 
cluster the selection rule prohibits transitions back
into the single-hole band and a different
band of many-body states with odd parity under reflection by
the $(1,1)$ axis is seen.
We have identified them as a
hole dressed by a cloud of spin defects
with intrinsic $d_{x^2-y^2}$ symmetry.
This band has almost no dispersion and
for all momenta accessible to our diagonalization
study remains $\sim$$2J$ above $E_f$, and thus is
unrelated to any low-energy physics.\\
While there is no well-defined Fermi surface in the bound states
we have studied, the obvious
validity of the rigid band approximation 
suggests that if a Fermi surface exists at all, 
it takes the form of small hole-pockets. The precise location
of these pockets in an infinite system
is impossible to predict on the basis of exact diagonalization
results; due to the near-degeneracy of the states
near the surface of the magnetic Brillouin zone,
the interaction between the holes is
decisive. Whereas the $16$ and $18$ site cluster
both suggest $(\pi,0)$ as the locus of the pockets,
this momentum seems to be largely unoccupied
by holes in the $20$-site cluster, where 
the largest shift from PES to IPES occurs at
$(2\pi/5,4\pi/5)$. The best one can say is that there
seems to be a trend for hole occupation at or near $(\pi,0)$.
Since it is the interaction between the holes,
which favours these $\vec{k}$-points,
the apparent contradiction with the
well known fact that the single hole ground state
has momentuma $(\pi/2,\pi/2)$ is not surprising.\\
Rigid band behaviour and
hole pockets in the $t$$-$$J$ model
are consistently suggested by a number of
previous exact diagonalization works.
Poilblanc and Dagotto\cite{PoilblancDagotto}
studied the $A(\bbox{k},\omega)$ for single hole states 
and concluded that the two-hole ground state in the
$4\times 4$ cluster shows
hole pockets at $(\pi,0)$, in agreement with
the present result. 
Stephan and Horsch\cite{StephanHorsch} studied $n(\bbox{k})$
and $A(\bbox{k},\omega)$ for the two-hole ground state
and concluded
on the contrary that there is neither rigid band behaviour
nor hole pockets. It should be noted, however, that whereas
Poilblanc and Dagotto employed a quantitative criterion
(presence or absence of a quasiparticle peak at the position of the
two-hole ground state energy), Stephan and Horsch based
their conclusions solely on the qualitative inspection
of a rather limited data set; as discussed above
(see Tab. \ref{table1}) $n(\bbox{k})$
is not reliable for assigning the Fermi surface and
our results for the spectral function in the $20$-site cluster
show for example that the Luttinger Fermi surface 
assigned there by Stephan and Horsch does not exist.
Next, Castillo and Balseiro\cite{CastilloBalseiro} 
computed the Hall constant and found its
sign near half-filling to be consistent with a hole-like
Fermi surface.
Gooding {\it et al.}\cite{Goodingetal} studied
the doping dependence of the spin correlation function
in clusters with special geometry and
also found indications of rigid-band
behaviour. Finally, a
study of $n(\bbox{k})$ in the range
$J>t$ (where the incoheren continua are negligible)
with an added density repulsion to preclude 
hole clustering\cite{EderOhta} shows unambiguous hole pockets.
It seems fair to say
that the available numerical results for the $t$$-$$J$ model,
when interpreted with care are all consistent with
rigid band behaviour and/or hole pockets.\\
In summary, we have performed a detailed study of the
doping dependence of the single particle spectral
function up to the largest clusters that are numerically
tractable. The results
show unambiguously that rigid-band behaviour is
realized in small clusters of $t$$-$$J$ model:
near the chemical potential, the main effect of the
doping consists in moving the Fermi energy into the `band'
of peaks observed at half-filling. Thereby the
parts of the quasiparticle band which
remain on the photoemisson side
are essentially unaffected,
the uppermost states of this band
simply cross the Fermi level and reappear
as the lowermost states of the inverse photoemission
spectrum. This behaviour is always realized, 
unless it is prohibited by a trivial reason, namely
a symmetry related selection rule.
In the latter case, there is no low energy IPES weight at all.
On the PES side, modifications of the spectral 
function which deviate from
the rigid band picture occur predominantly
at energies far from
$E_f$, and hence should be unrelated to any
Fermi surface physics. In particular, the gains and losses of
PES weight which lead to the formation
of a `large Fermi surface' in the momentum distribution
upon doping originate from
addition and depletion of incoherent weight
deep below the Fermi energy.
On the IPES side, there is no indication
of the emerging of `new states' at low energies
in the course of doping. Such new states in the IPES
spectrum are seen
only at high energies, and hence also 
should be unrelated to any low energy physics.
Available exact diagonalization data
are all in all consistent with this interpretation.\\
Then, we are left with the problem to reconcile 
the emerging picture for the $t$$-$$J$ model with
experiments on high-temperature superconductors.
While some transport properties
are quite consistent with a rigid-band/hole pocket 
scenario\cite{Trugman},
the main problem is with angular resolved photoemission
experiments\cite{King}. These show peaks 
which disperse towards
the Fermi energy and vanish there, as one would expect
for a band crossing. Thereby the locus of the
`crossing points' is remarkably consistent with
the predictions of band theory, which in turn is
inconsistent with hole pockets.
In a Fermi liquid, the contradicting 
quantities actually fall into two distinct classes: 
photoemission spectra depend on 
the wave function renormalization constant $Z_h$,
transport properties do not. Hence, if one wants to resolve the
discrepancy concerning the volume of the Fermi surface
entirely within a Fermi liquid-like picture,
the simplest way would be to assume a `small' Fermi surface
(to model the transport properties)
and explain the photoemission results by a
systematic variation of $Z_h$
along the band which forms the Fermi surface,
i.e.  similar to the `shadow band' 
picture\cite{KampfSchrieffer}.
A trivial argument for such a strong variation in
$Z_h$ would be that
irrespective of the actual band structure, a distribution of
PES weight in the Brillouin zone that resembles the
nointeracting (band theory)
Fermi surface always optimizes the
expectation value of the kinetic energy.
A wave function which gives a substantially different
distribution of spectral weight consequently has a very
unfavourable kinetic energy and hence is ruled as the
ground state from the very beginning.
Also, it should be noted that the spectral 
weight in the shadow band
should even decrease when the charge fluctuations in the
original Hubbard model are taken into account: 
when going back to the Hubbard model,
the (negative) exchange energy
in the $t$$-$$J$ model then is split into a
(positive) contribution from the Hubbard repulsion and a
(negative) gain in kinetic energy, so that the total expectation
value of the kinetic energy certainly becomes more negative
than for the $t$$-$$J$ model.
Via the kinetic-energy sum rule (\ref{kinsum}),
one can infer that this necessarily leads to an even more 
free-electron like distribution of spectral weight.
All in all, adopting the rigid-band behaviour found above, the
available diagonalization data for the $t$$-$$J$ 
model
\cite{StephanHorsch,PoilblancDagotto,CastilloBalseiro,Goodingetal}
as a whole then are reasonably consistent with the above scenario.\\
 
It is a pleasure for us to acknowledge numerous instructive 
discussions with Professor S. Maekawa.
Financial support of R. E.
by the Japan Society for the Promotion of
Science is most gratefully acknowledged.

\begin{table}
\caption{Momentum distribution in the ground state of
of the $t$$-$$J_z$ model ($t/J_z=2$) in a staggered 
magnetic field ($0.1J_z$) and the
$t$$-$$J$ model ($t/J$$=$$2$) in a $4$$\times$$4$ cluster.
To stabilize a ground state with the same point group
symmetry ($B_1$) as for the $t$$-$$J$ model, a $2^{nd}$ nearest
neighbor hopping term of strength $-$$J_z/10$ has been added to the
$t$$-$$J_z$ model.}
\begin{tabular}{c | c c c c c c}
$\bbox{k}$ &$(0,0)$ & $(\frac{\pi}{2},0)$ & $(\pi,0)$ 
& $(\frac{\pi}{2},\frac{\pi}{2})$ &
$(\frac{\pi}{2},\pi)$ & $(\pi,\pi)$ \\
\hline
$n(\bbox{k})(t-J_z)$ 
& 0.5481  & 0.5374 & 0.3491 & 0.4993 & 0.3387 & 0.2522\\
$n(\bbox{k})(t-J)$ 
& 0.5565  & 0.5421 & 0.3378 & 0.4974 & 0.3225 & 0.3197\\
\end{tabular}
\label{table1}
\end{table}
\figure{Comparison of the photoemsission
        spectra at half-filling
        and in the ground state with 2 holes:
        $A_{0,-}(\bbox{k},-\omega)$ (dotted line)
        $A_{2,-}(\bbox{k},-\omega)$ (full line)
        $A_{2,+}(\bbox{k},\omega)$ (dashed-dotted line)
        for the $16$ and $18$-site cluster.
        $\delta$ functions have been replaced by Lorentzians
        of width $0.1t$.
\label{spec1} }
\figure{Same as Fig. \ref{spec1} but for the $20$-site cluster.
\label{spec21} }
\figure{PES spectrum $A_{0,-}(\bbox{k},\omega)$ (full line)
        and IPES spectrum $A_{2,+}(\bbox{k},\omega)$ 
        (dashed-dotted line) for the $16$ and $18$-site cluster.
        The reference energies are identical so that 
        direct comparison of the peaks is possible,
        $\delta$ functions have been replaced by Lorentzians
        of width $0.05t$.
\label{spec2} }
\figure{Same as Fig. \ref{spec2} but for the $20$-site cluster.
\label{spec22} }
\figure{PES spectrum $\tilde{A}_{0,-}(\bbox{k},\omega)$ (full line)
        and IPES spectrum $A_{2,+}(\bbox{k},\omega)$ 
        (dashed-dotted line) for momenta along $(1,1)$
        in the $16$ and $18$-site cluster.
        The reference energies
        are identical so that direct comparison of the peaks
        is possible,
        $\delta$ functions have been replaced by Lorentzians
        of width $0.05t$.
\label{spec3} }
\figure{Comparison of the $t/J$-dependence of the
        PES pole strength at $(\pi/2,\pi/2)$ at half-filling
        (squares) and in the two-hole ground state (triangles)
        and the added weights of the lowest IPES peak and 
        highest PES peak
        at $(\pi,0)$ in the two-hole ground state (circles) 
\label{poles} }
\figure{Schematic quasiparticle band structure in the 
        neighborhood of $E_f$ for the $16$ and $18$-site cluster.
        Up-triangles (squares) give the position of the highest peak
        in $A_{2,-}(\bbox{k},-\omega)$ 
        down triangles (circles) the position of the lowest peak
        in $A_{2,+}(\bbox{k},\omega)$ 
        for the $4$$\times$$4$ ($18$-site) cluster.
        The positions of the highest peaks in 
        $A_{0,-}(\bbox{k},-\omega)$ (dots) are also given.
        The various groups of spectra have been shifted so that
        the energies of the respective PES peak at $(0,0)$ coincide
        (shift between doped $16$ and $18$-site cluster: $0.275t$).
\label{band} }
                                                        
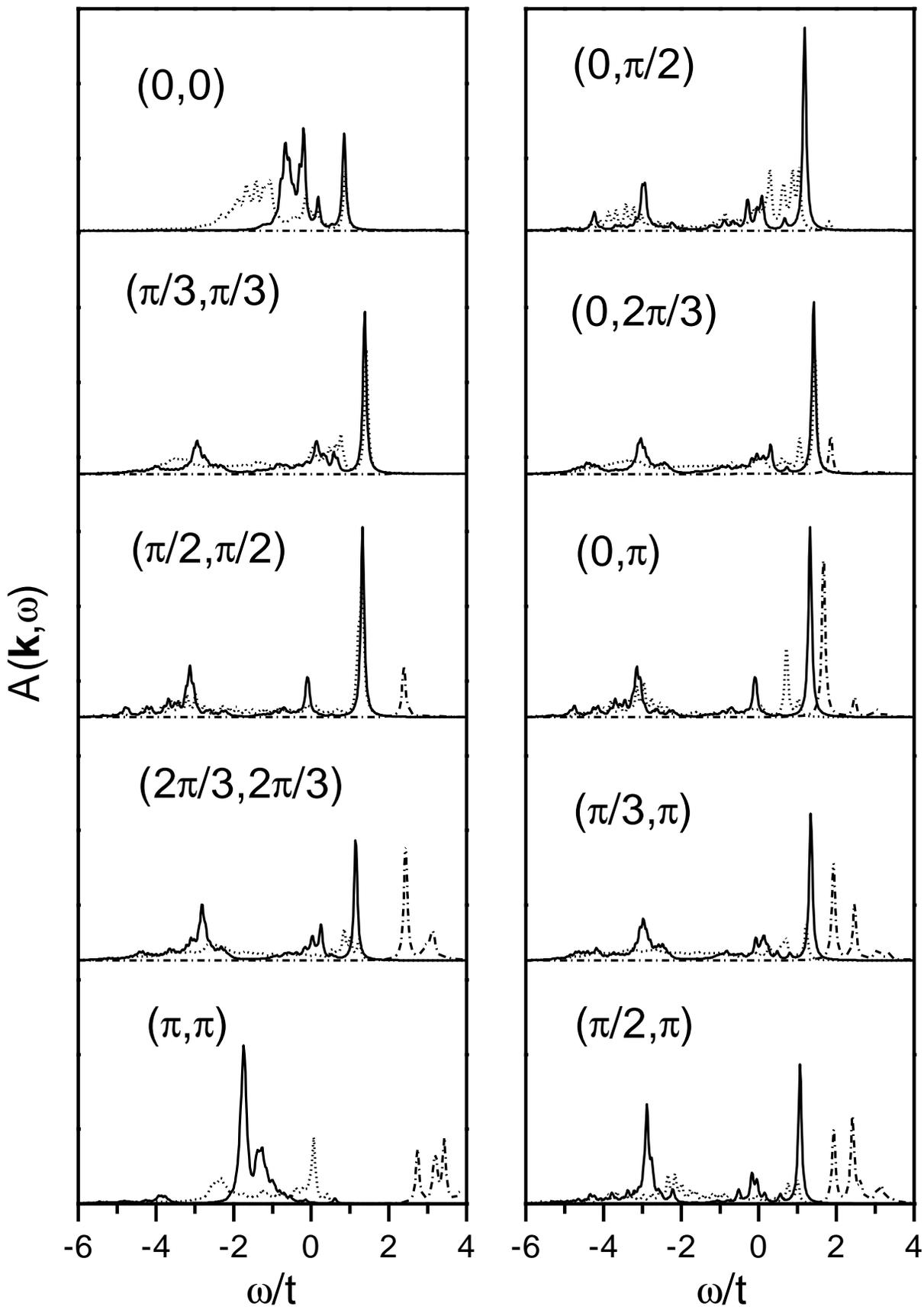
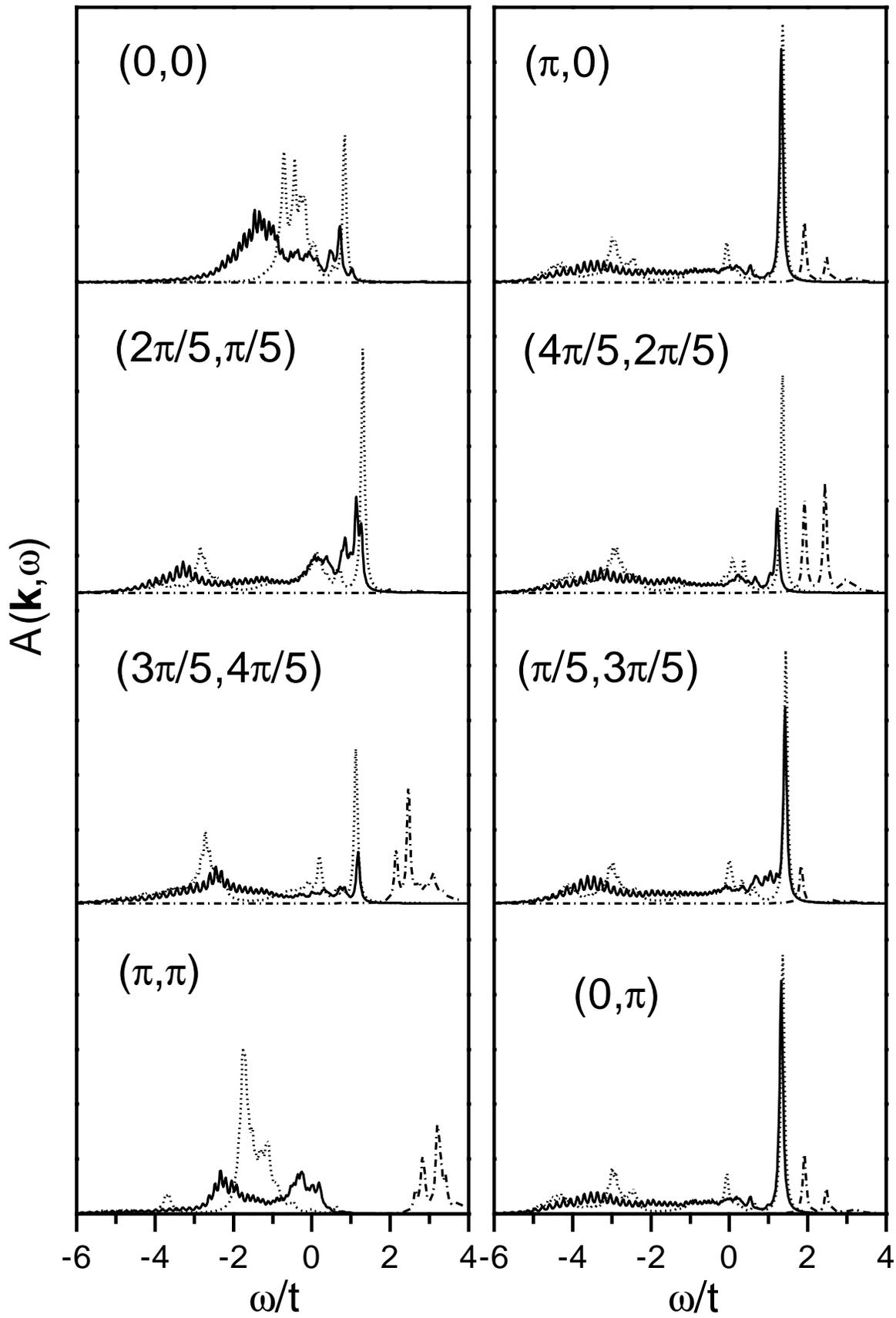
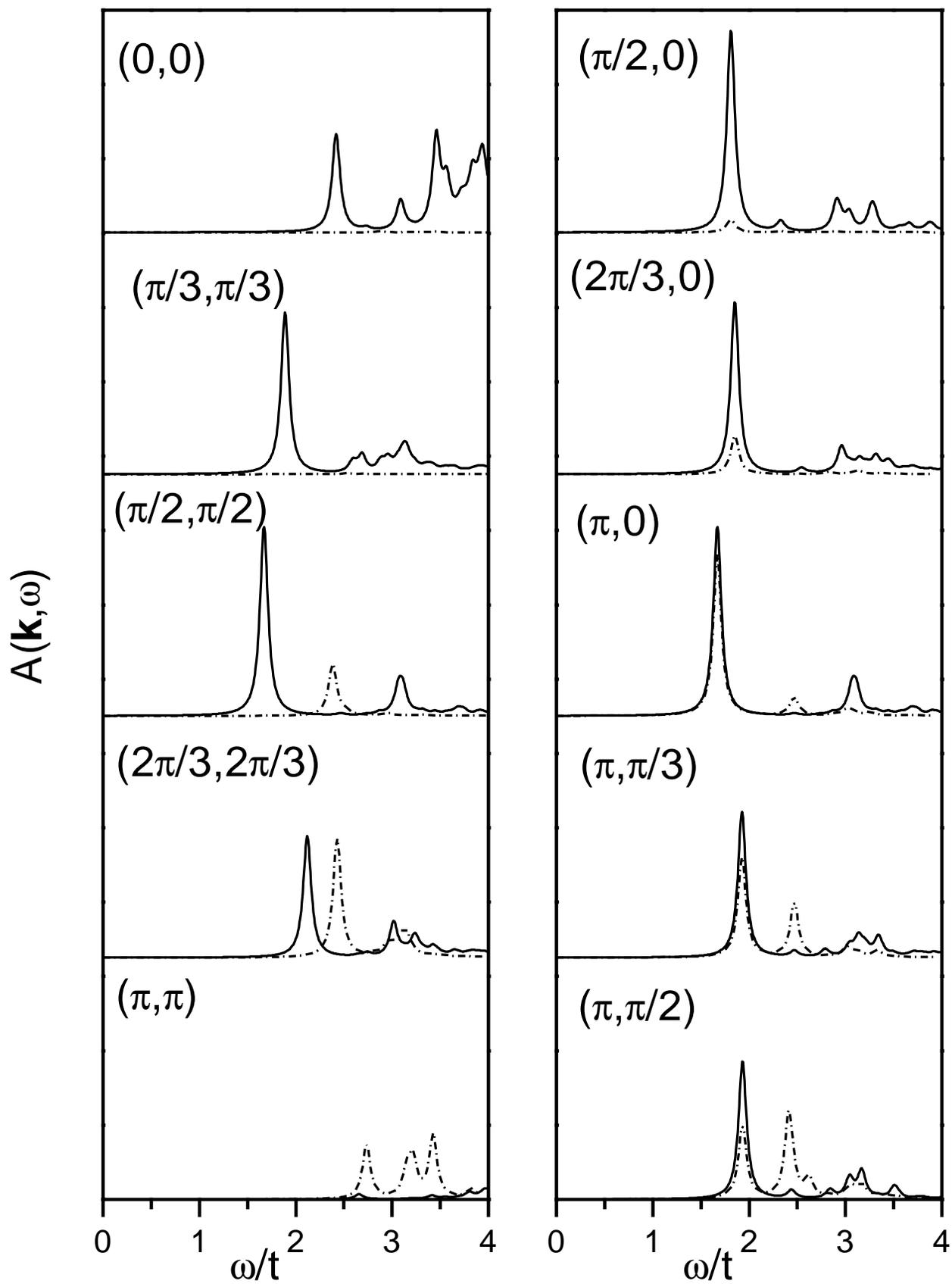
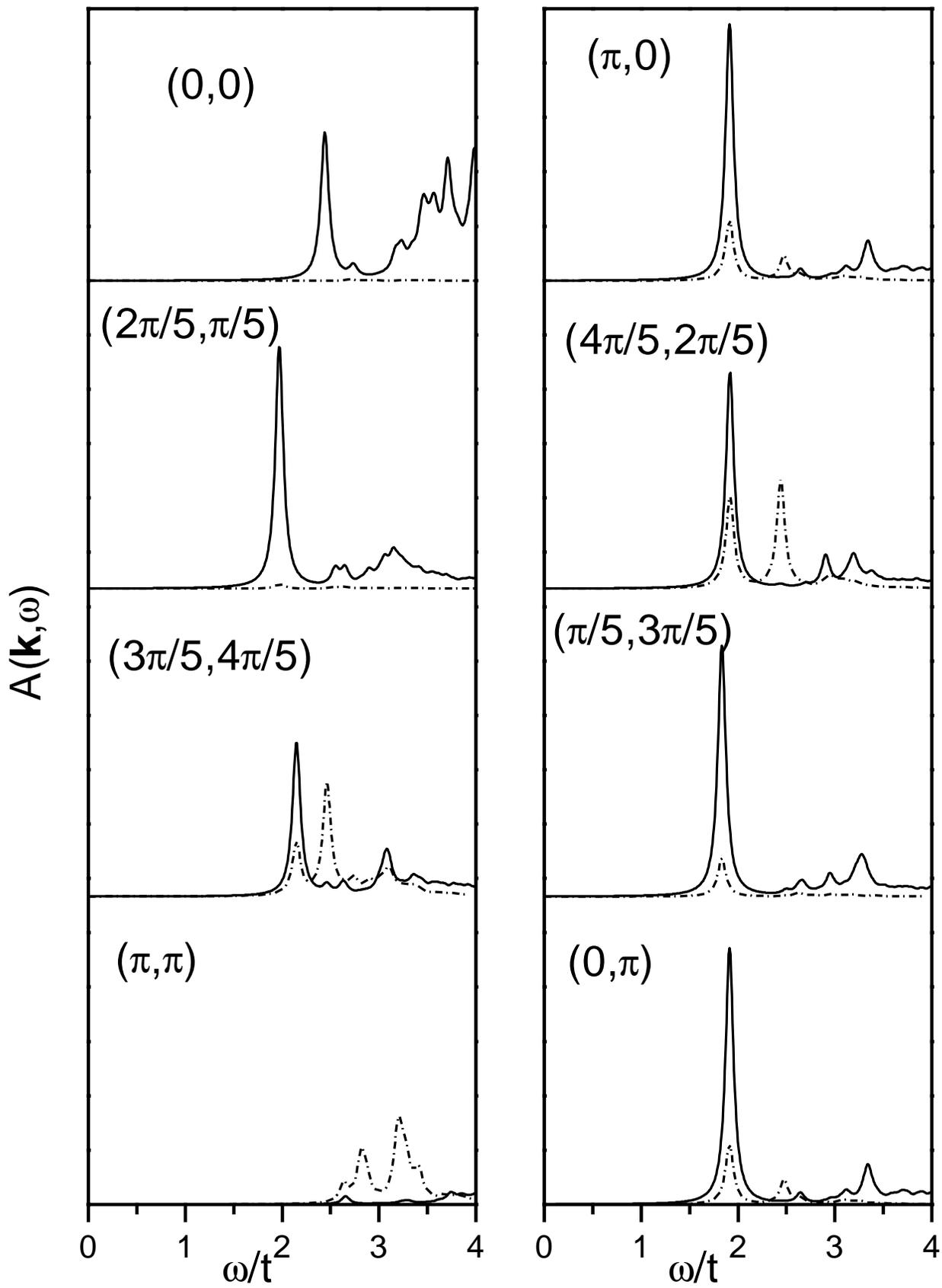
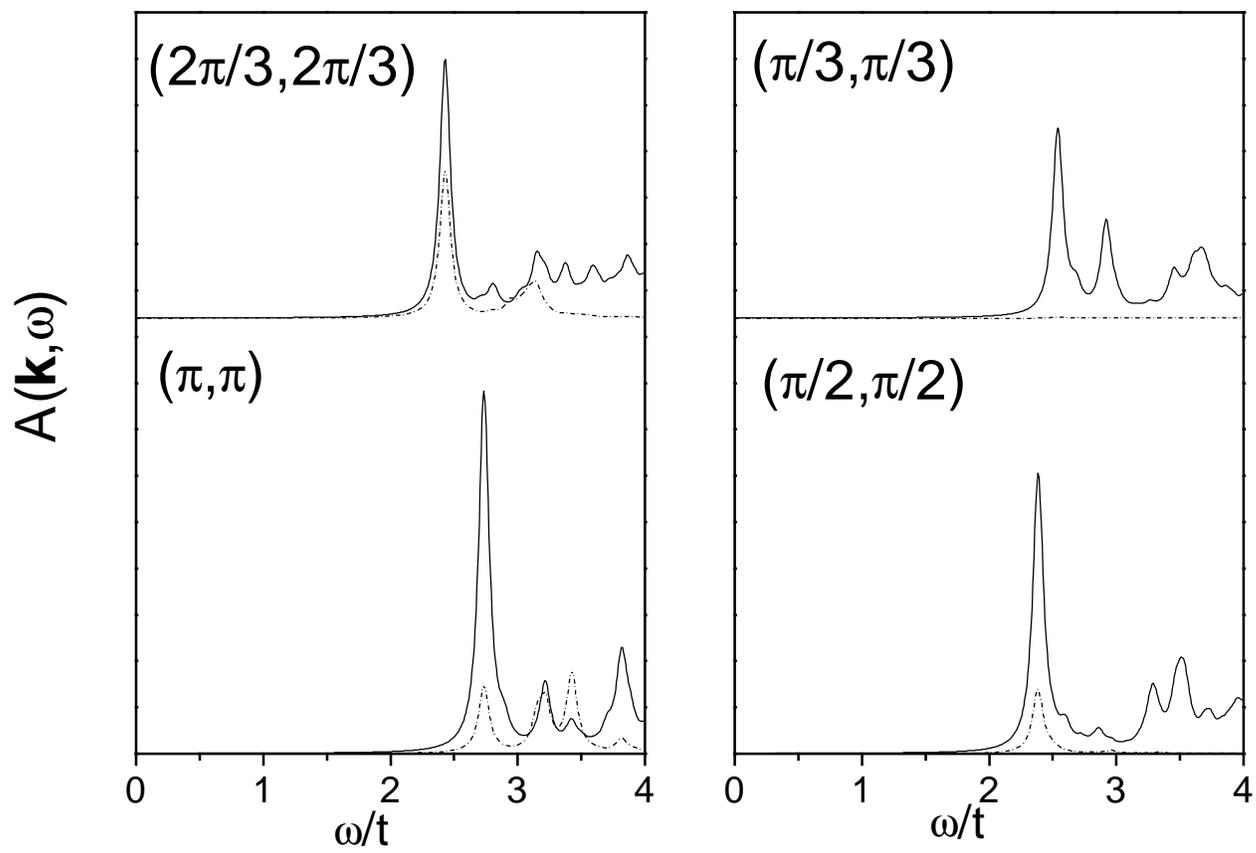
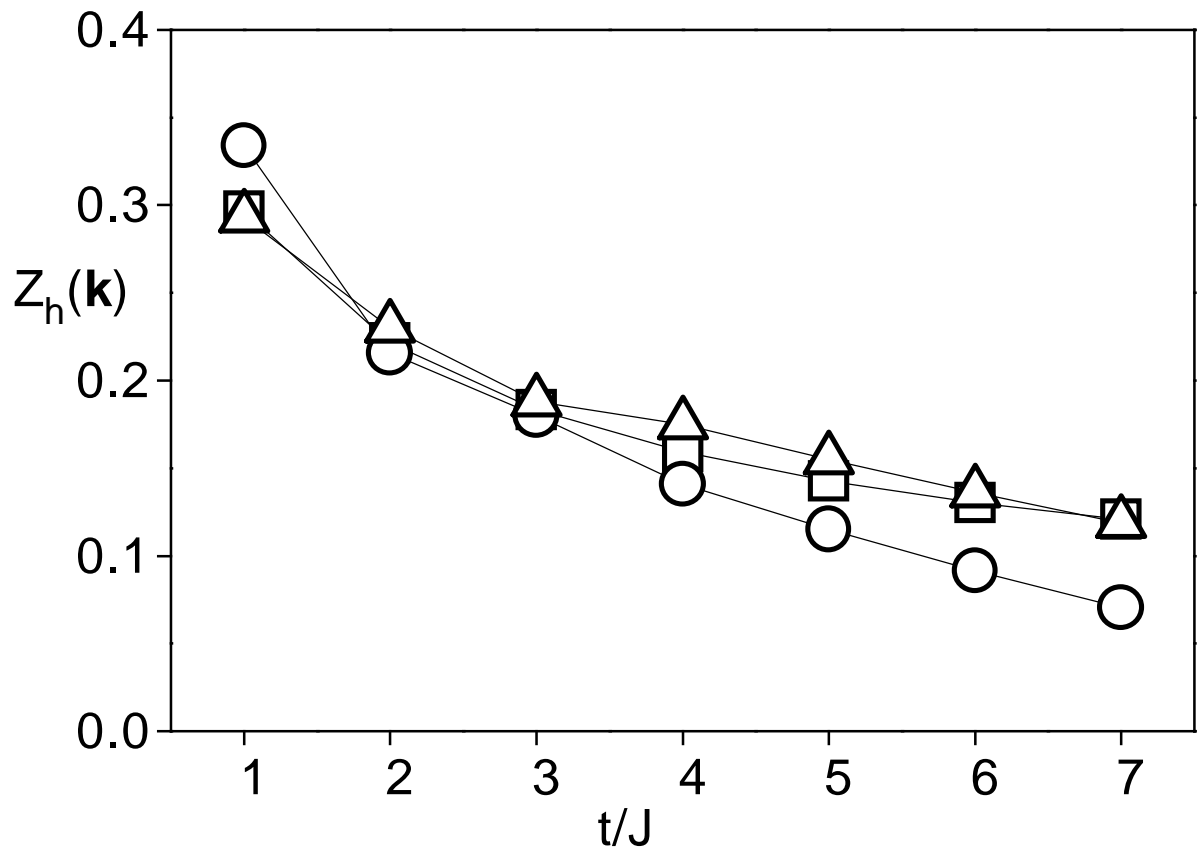
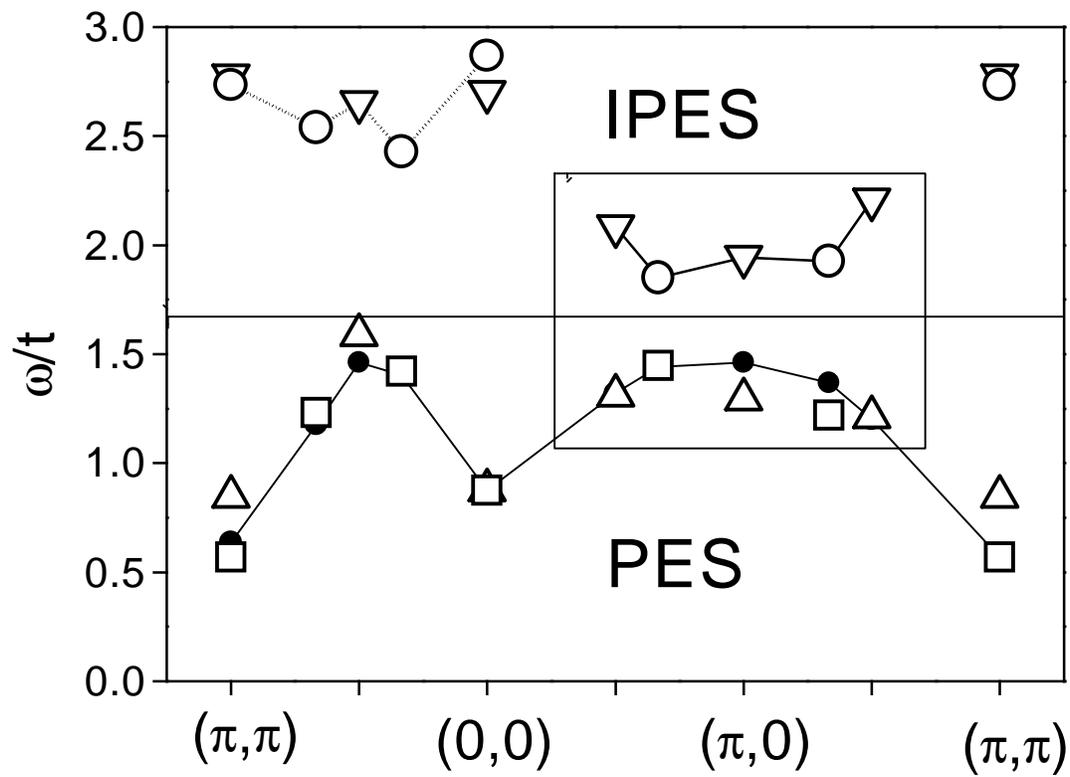
\end{document}